%% file: main.tex
\documentclass[conference,a4paper]{IEEEtran}
\usepackage[left=1.62cm,right=1.62cm,bottom=4.16cm,top=1.9cm]{geometry}
  	\usepackage[pdftex]{graphicx}
  	\graphicspath{{../pdf/}{../jpeg/}}
	\DeclareGraphicsExtensions{.pdf,.jpeg,.png}
    \usepackage{balance}
    \usepackage{titlesec}
    \usepackage{pdfpages}
    \usepackage{amssymb}
    \usepackage{color}
    \usepackage{soul}
	\usepackage[cmex10]{amsmath}
	\usepackage{mathabx}
	\usepackage{array}
	\usepackage{mdwmath}
	\usepackage{mdwtab}
	\usepackage{eqparbox}
	\usepackage{url}
    \usepackage{ragged2e}
    \usepackage{fancyhdr}
    \usepackage{tabularx}
    \usepackage{float}
    \usepackage{multicol}
    \usepackage{listings}
    \usepackage[breakable]{tcolorbox}
 \usepackage{tabularx}
\usepackage{float}
\usepackage{multicol}
\usepackage{listings}
\usepackage{subcaption}
\usepackage{algpseudocode}
\usepackage[breakable]{tcolorbox}
\usepackage{afterpage}
\usepackage{placeins}
\usepackage{makecell}
\usepackage[utf8]{inputenc}
\usepackage{amsmath}
\usepackage{algorithm}
\usepackage{algpseudocode}
 
\fancypagestyle{titlepage}{
\fancyhf{}

\fancyfoot[C]{\thepage}
}

\title{Enhanced Real-Time Threat Detection in 5G Networks: A Self-Attention RNN Autoencoder Approach for Spectral Intrusion Analysis}

\author{\IEEEauthorblockN{Mohammadreza Kouchaki, Minglong Zhang, Aly S. Abdalla \\ Guangchen Lan\IEEEauthorrefmark{2}, Christopher G. Brinton\IEEEauthorrefmark{2}, Vuk Marojevic}
\IEEEauthorblockA{Department of Electrical and Computer Engineering\\
Mississippi State University, USA\\
\IEEEauthorblockA{\IEEEauthorrefmark{2}Purdue University, USA\\
Email: \{mk1682, mz354, asa298, vm602\}@msstate.edu} \IEEEauthorrefmark{2}\{lan44, cgb\}@purdue.edu}
}

\date{May 12 2023}

\begin{document}

\maketitle
\begin{abstract}
In the rapidly evolving landscape of 5G technology, safeguarding Radio Frequency (RF) environments against sophisticated intrusions is paramount, especially in dynamic spectrum access and management. This paper presents an enhanced experimental model that integrates a self-attention mechanism with a Recurrent Neural Network (RNN)-based autoencoder for the detection of anomalous spectral activities in 5G networks at the waveform level. Our approach, grounded in time-series analysis, processes in-phase and quadrature (I/Q) samples to identify irregularities that could indicate potential jamming attacks. The model's architecture, augmented with a self-attention layer, extends the capabilities of RNN autoencoders, enabling a more nuanced understanding of temporal dependencies and contextual relationships within the RF spectrum.
Utilizing a simulated 5G Radio Access Network (RAN) test-bed constructed with srsRAN 5G and Software Defined Radios (SDRs), we generated a comprehensive stream of data that reflects real-world RF spectrum conditions and attack scenarios. The model is trained to reconstruct standard signal behavior, establishing a normative baseline against which deviations, indicative of security threats, are identified. The proposed architecture is designed to balance between detection precision and computational efficiency, so the LSTM network, enriched with self-attention, continues to optimize for minimal execution latency and power consumption. Conducted on a real-world SDR-based testbed, our results demonstrate the model's improved performance and accuracy in threat detection.

\end{abstract}
\section{Introduction}

The advent of the fifth generation (5G) of wireless communication systems has ushered in an unprecedented era of connectivity and innovation. With its promise of higher data rates, reduced latency, and increased capacity, 5G is set to revolutionize various sectors, including smart cities, autonomous vehicles, and the Internet of Things (IoT) \cite{10008713}. Meanwhile, on the other side, it introduces significant security vulnerabilities, particularly in radio frequency (RF) communications, caused by high-density networks with a large number of access points and user equipment (UE). The flexible allocation of spectrum in 5G renders the monitoring and securing channel access more complicated \cite{Sagduyu2023Security}. Among diverse threats, RF jamming attacks emerge as a substantial threat, undermining the reliability and functionality of critical 5G network services that are fundamental to sectors like IoT and autonomous vehicles. For instance, remote surgery and autonomous driving \cite{Aldabbagh2021vehicle}, require ultra-reliable low-latency communications, which can be compromised by malicious interference. Unfortunately, traditional network security mechanisms cannot effectively cope with the threats due to multiple factors. The difficulties for traditional methods may include new evolved threats; highly dynamic spectrum access and large-scale networks. Therefore, the variety and sophistication of potential attacks, such as advanced persistent threats (APTs) and intelligent jamming, necessitate more advanced detection mechanisms.

In fact, taking into account the traditional intrusion detection systems (IDS) in the context of 5G's unique demands and threat landscape, the potential challenges to address the RF jamming attacks are detailed as follows. \textbf{Dynamic spectrum access} in 5G networks introduces a layer of complexity where the 5G's spectrum is characterized by its fluidity – bandwidths vary, access policies shift frequently, and modulation schemes adapt in real-time \cite{Ramakrishnan2022Dynamic}. This presents a significant challenge for traditional IDS, which are typically engineered for more static environments. The \textbf{evolving threat landscape} in 5G networks incorporate advanced threats like adaptive jamming and complex advanced persistent threats (APTs)~\cite{9542973}. 
These modern attacks often do not follow repetitive patterns and are designed to adapt to countermeasures, making them particularly challenging to detect by traditional IDS. The exponential growth in the number of connected devices and network nodes within 5G networks, coupled with the openness and flexibility introduced by the adoption of Open Radio Access Network (O-RAN) architecture, exacerbates these challenges. The open interfaces and disaggregated components in O-RAN can be exploited by attackers, creating vulnerabilities that traditional security methods are not equipped to handle effectively \cite{10520642}.

To effectively counter the aforementioned challenges in 5G networks, an advanced intrusion detection system (IDS) is paramount. This IDS must be both reactive, to counter known threats, and proactive, to adapt to emerging, unseen attack patterns. Critically, it must achieve this balance while being scalable and resource-efficient, ensuring that the intrinsic performance benefits of 5G are not compromised as the network expands in size and complexity.

Addressing these requirements, our research proposes a novel IDS framework that synergizes an efficient self-attention mechanism with a recurrent neural network (RNN)-based autoencoder. This combination is strategically chosen to tackle the unique challenges posed by 5G networks. The self-attention mechanism of our solution enables the IDS to adaptively focus on specific spectrum parts more prone to anomalies \cite{Vaswani2017AttentionIA} to enhance its efficacy in safeguarding against spectrum-related vulnerabilities. The integration of unsupervised learning capabilities in the self-attention-equipped RNN autoencoder enables the detection of both known and novel attack patterns. By learning complex dependencies within the data, the model is equipped to identify emerging cyber threats that were not part of its initial training set. The self-attention mechanism computational efficiency translates into the ability for parallel processing, a crucial factor in reducing the computational load. This ensures that the IDS can keep pace with the growing size and complexity of 5G networks, offering robust threat detection without significant resource overheads.

Our research presents a sophisticated approach to addressing the complex challenges associated with RF intrusion detection in 5G networks. We have developed a model that effectively combines the temporal processing capabilities of Recurrent Neural Networks (RNNs) with the contextual sensitivity afforded by self-attention mechanisms. This integration results in a robust, efficient, and scalable system. Specifically designed to process and analyze the time-series data characteristic of the RF spectrum, our model excels at identifying anomalies indicative of potential jamming attacks.

A notable aspect of our work is the deployment of a 5G Radio Access Network (RAN) test-bed, along with comprehensive databases, to facilitate the training and inference phases of our model. We have structured a sequential two-part methodology focusing on anomaly detection and subsequent classification. This approach is further bolstered by rigorous experimental validation and an extensive analysis of performance metrics. Our model demonstrates a high proficiency in detecting a diverse range of cyber threats, positioning it as a viable and effective tool for practical application in real-world 5G scenarios.

The ensuing sections of this paper will explore the background, methodology, experimental setup, results, and provide a detailed analysis of our proposed model. This comprehensive examination will underscore the model’s effectiveness in protecting 5G networks against sophisticated and evolving RF threats, thus showcasing its potential as a critical asset in modern network security frameworks.

\section{Related Work}

Intrusion detection systems (IDS) are crucial for identifying unauthorized access or anomalies in network traffic, which could indicate a cyber-attack. Traditional IDS tools have been designed to protect network perimeters by monitoring traffic and identifying known attack patterns. However, the dynamic and complex nature of 5G networks requires more adaptive and intelligent IDS solutions. The integration of machine learning techniques with IDS provides the ability to learn from the network behavior and predict potential threats based on anomaly detection \cite{buczak2016survey, bank2023autoencoders}. Recurrent Neural Networks (RNNs) are particularly well-suited for this task due to their ability to process sequences of data and retain information over time, making them ideal for analyzing time-dependent network traffic \cite{yin2017deep}.

The use of autoencoders in IDS offers a promising approach to anomaly detection. An autoencoder is a type of neural network that is trained to copy its input to its output, with the goal of learning a representation (encoding) for the data. In the context of intrusion detection, autoencoders can learn to reconstruct normal network traffic and, by measuring the reconstruction error, can identify anomalies that deviate from the learned pattern \cite{sakurada2014anomaly}. This method is particularly effective in detecting novel or sophisticated attacks that do not match any known signature. The application of autoencoders in 5G networks is especially pertinent due to the high volume and velocity of data, where traditional signature-based IDS tools may fail to keep pace with the evolving threat landscape \cite{lam2020machine}. 

The deployment of RNN-based autoencoders in 5G networks addresses several challenges associated with the high-speed, high-volume, and low-latency requirements of modern wireless systems. 5G networks are characterized by their use of a diverse spectrum, including millimeter-wave frequencies, which necessitates a more nuanced approach to spectrum monitoring and security \cite{ahmad2019security}. The ability of RNNs to process sequential and time-series data makes them particularly useful for detecting irregularities in the RF spectrum that could indicate jamming or other forms of RF interference. Furthermore, the distributed nature of 5G architecture, with its emphasis on edge computing, calls for distributed IDS solutions that can operate effectively at the network's edge, closer to the user equipment (UE) \cite{wang2019edge, lan2023edge}.
\begin{figure*}[t] 
\centering
\includegraphics[width=0.91\textwidth]{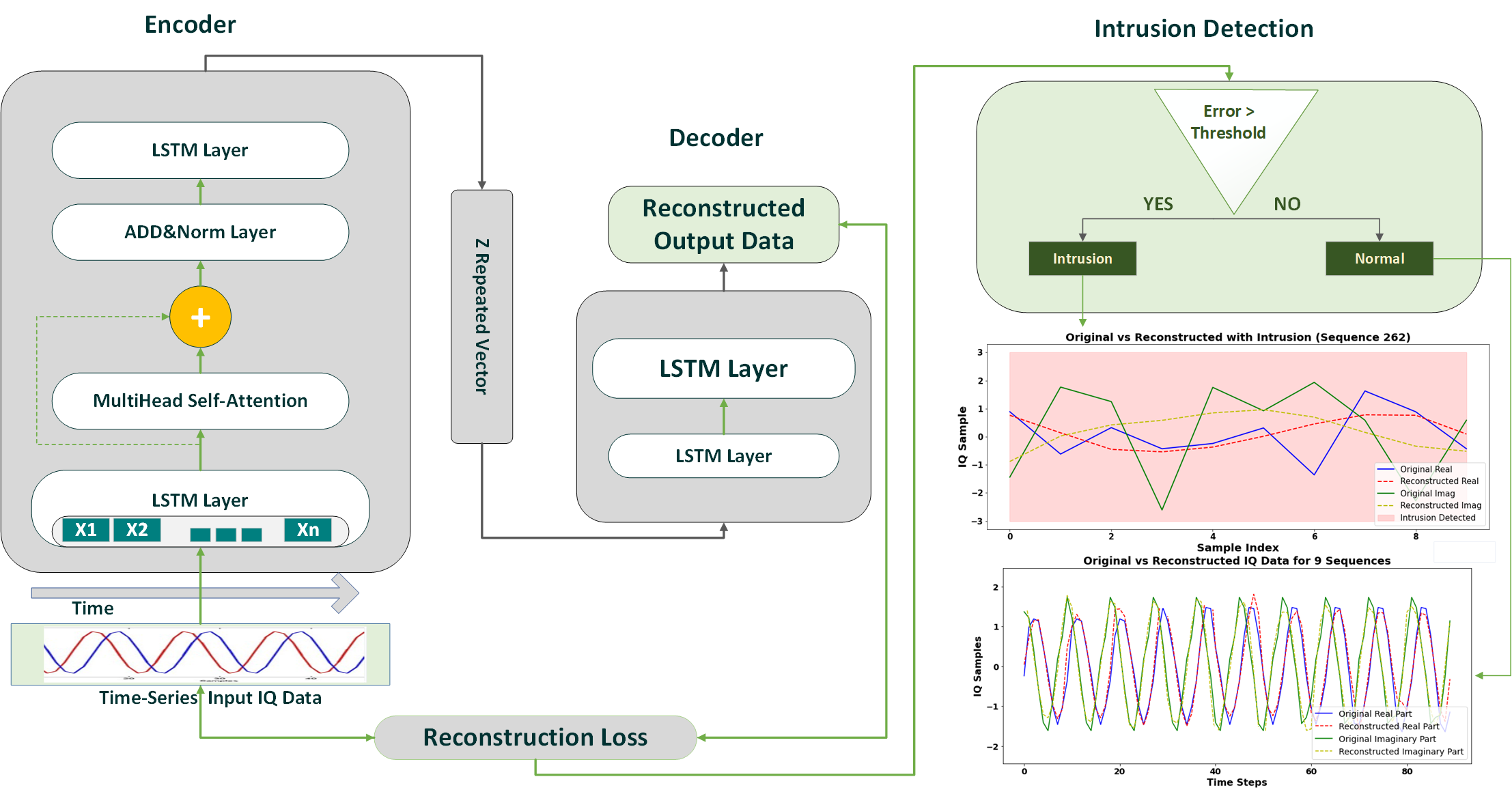}
\caption{Proposed Self-Attention RNN-Based Autoencoder.}
\label{arch1}
\vspace{-10pt}
\end{figure*}
Autoencoders and attention mechanisms have seen extensive application and development across various fields, evolving significantly in their design and utility. A comprehensive review \cite{LI2023110176} delves into the theory, different models, and diverse applications of self-coding neural networks, highlighting their versatility in fields like image classification and natural language processing.

The study \cite{chaudhari2021attentive} offers a structured overview of attention models, underlining their importance in enhancing model performance, particularly in classification tasks. Moreover, advancements in self-attention mechanisms, as explored in \cite{article}, have shown remarkable improvements in classifier efficiency, further augmented by techniques like Generative Adversarial Networks (GANs). Attention mechanisms also play a critical role in intelligent fault diagnosis, as demonstrated in \cite{LV2022111594}, who categorizes these mechanisms to facilitate better understanding and application in various domains. Lastly, the survey on masked autoencoders \cite{zhang2022survey} sheds light on the promising direction of self-supervised learning, revealing new insights and applications for these models.

\section{Methodology}

\subsection{Theoretical Framework}
The proposed methodology utilizes an autoencoder based on a recurrent neural network (RNN), enhanced with a multihead attention layer, for time-series anomaly detection in 5G networks. This advanced autoencoder is specifically crafted to recognize the IQ data patterns of the network and identify deviations that may signal potential intrusions or malfunctions. The complete architecture is illustrated in Figure \ref{arch1}. We will proceed to dissect and discuss this architecture, focusing on its key components.

\subsubsection{RNN Autoencoder Architecture}
An autoencoder is a type of artificial neural network used to learn efficient representations of input data, called encodings, through an unsupervised learning process. It consists of two main components: an encoder and a decoder. The encoder compresses the input into a latent-space representation, and the decoder reconstructs the input data from this representation. RNNs are a class of neural networks that exhibit temporal dynamic behavior for a time sequence. Unlike feedforward neural networks, RNNs have recursive connections, allowing them to maintain a 'memory' of previous inputs in their internal state. This feature makes RNNs particularly well-suited for modeling time-series data. The RNN-based autoencoder model can be mathematically described as follows:

\textbf{Encoder:} The encoder part of the RNN processes the input sequence $x = (x_1, x_2, ..., x_T)$, where $x_t \in \mathbb {R}^n$ is the input at time step $t$. The hidden state $h_t$ of the RNN at time $t$ is updated by:
\vspace{-7 pt}
\begin{equation}
h_t = f(W_{hh}h_{t-1} + W_{hx}x_t + b_h)
\vspace{-5 pt}
\end{equation}
where \(W_{hh}\) and \(W_{hx}\) are the weight matrices, \(b_h\) is the bias vector, and \(f\) is a nonlinear activation function, typically a sigmoid or tanh function.

\textbf{Decoder:} The decoder aims to reconstruct the input sequence from the hidden state. The output \(\hat{x}_t\) at time step \(t\) is given by:
\vspace{-7 pt}
\begin{equation}
\hat{x}_t = g(W_{xh}h_t + b_x)
\end{equation}
where \(W_{xh}\) is the weight matrix, \(b_x\) is the bias vector for the decoder, and \(g\) is the activation function, which may differ from \(f\) depending on the specific requirements of the input data reconstruction.

\subsection{Self-Attention in RNN Autoencoder}

The self-attention mechanism is a transformative approach in neural networks that allows different parts of the input sequence to be weighed differently in terms of their importance or relevance. It is particularly effective in identifying intricate patterns and dependencies in sequential data, such as 5G network traffic. By integrating self-attention into the RNN autoencoder, particularly in the encoder, the model gains an enhanced capacity to focus on significant features of the input sequence, thereby improving anomaly detection capabilities. The whole process, as outlined in the pseudocode of Algorithm \ref{alg1}, encompasses our proposed algorithm. In the following sections, we will delve into a detailed discussion of this process.

The mathematical framework of the self-attention mechanism as integrated into our RNN autoencoder is detailed through a series of equations that delineate the computation of alignment scores, attention weights, and the resultant context vector \cite{Vaswani2017AttentionIA}.
\begin{algorithm}[H]
\footnotesize
\caption{Self-Attention Enhanced LSTM-based Autoencoder for Intrusion Detection in Time-Series Data}
\begin{algorithmic}[1]
\Statex \textbf{Input:} Time-series sequence data $X = \{x_1, x_2, \dots, x_T\}$, where $x_t \in \mathbb{R}^n$
\Statex \textbf{Output:} Intrusion detection model $M$
\Statex
\Procedure{EnhancedAutoencoder}{}
    \State Initialize parameters $\Theta$ for LSTM and self-attention layers
    \For{each epoch}
        \For{each sequence $x$ in $X$}
            \State $\tilde{x} \gets \Call{CorruptSequence}{x}$
            \State $h_0 \gets \Call{InitializeHiddenState}{}$
            \For{$t \gets 1$ to $T$}
                \State $h_t \gets \Call{LSTMCell}{\tilde{x}_t, h_{t-1}, \Theta}$
                \State $h'_t \gets \Call{SelfAttention}{h_t, \Theta}$
            \EndFor
            \State $\hat{x} \gets \Call{Reconstruct}{h', \Theta}$
            \State $L \gets \Call{Loss}{x, \hat{x}}$
            \State Update $\Theta$ using gradients of $L$
        \EndFor
    \EndFor
    \State \Return $\Theta$
\EndProcedure
\Statex
\Function{SelfAttention}{$h, \Theta$}
    \State \textbf{bold} $Q, K, V \gets h\Theta^Q, h\Theta^K, h\Theta^V$
    \State $A \gets \Call{Softmax}{\frac{QK^T}{\sqrt{d_k}}}V$
    \State \Return $A$
\EndFunction
\Statex
\Function{Reconstruct}{$h', \Theta$}
    \For{$t \gets T$ down to $1$}
        \State $h'_t \gets \Call{LSTMCell}{h'_t, \Theta}$
    \EndFor
    \State $\hat{x} \gets h'_1\Theta^x$
    \State \Return $\hat{x}$
\EndFunction
\Statex
\Function{Loss}{$x, \hat{x}$}
    \State $L \gets \frac{1}{T}\sum_{t=1}^{T}(x_t - \hat{x}_t)^2$
    \State \Return $L$
\EndFunction
\end{algorithmic}
\label{alg1}
\end{algorithm}
\vspace{-8 pt}

\paragraph{Alignment Score Calculation:}
The alignment score is a crucial metric that represents the degree of relevance between different positions within the input sequence. It is computed as follows:
\vspace{-5 pt}
\begin{equation}
e_{tj} = a(h_t, h_j)
\vspace{-5 pt}
\end{equation}
where \( e_{tj} \) denotes the alignment score between the hidden state at the current time step \( h_t \) and the hidden state at another time step \( h_j \). The function \( a \) can be a simple feedforward neural network or any other suitable scoring function.

\paragraph{Attention Weights:}
The attention weights are obtained by normalizing the alignment scores using the softmax function, as shown below:
\vspace{-5 pt}
\begin{equation}
\alpha_{tj} = \frac{\exp(e_{tj})}{\sum_{k=1}^{T} \exp(e_{tk})}
\vspace{-5 pt}
\end{equation}
Here, \( \alpha_{tj} \) signifies the normalized importance of the hidden state \( h_j \) in the context of the hidden state \( h_t \).

\paragraph{Context Vector:}
The context vector is a pivotal element that encapsulates a weighted summary of the input sequence, and is computed by:
\vspace{-5 pt}
\begin{equation}
c_t = \sum_{j=1}^{T} \alpha_{tj}h_j
\vspace{-5 pt}
\end{equation}
The vector \( c_t \) at any given time step \( t \) is a weighted aggregation of all hidden states, modulated by the attention weights. It forms an enhanced representation of the input sequence by emphasizing the most pertinent segments.

In our proposed model, the self-attention mechanism is seamlessly integrated into the encoder of the RNN autoencoder. This integration modifies the encoder's computational processes to incorporate the context vector \( c_t \), thereby reinforcing the model's acuity in homing in on salient features within the 5G network traffic data. Such focused attention is instrumental in elevating the detection of anomalies by accentuating the features that are most indicative of potential intrusions or malfunctions.

\subsubsection{Enhanced Encoder with Self-Attention}
The encoder within the RNN autoencoder is augmented with a self-attention layer as follows:
\begin{itemize}
    \item The hidden states \( h_t \) are first generated by the LSTM cell for each time step.
    \item The self-attention mechanism computes a context vector \( c_t \) for each time step, which is then used to enrich the hidden states.
    \item The attention-augmented hidden states \( h'_t \) are further processed to reconstruct the input sequence, aiming to minimize the discrepancy between the original and the reconstructed sequence.
\end{itemize}

The integration of self-attention yields several benefits, including improved feature representation by capturing complex dependencies within the data, enhanced anomaly detection sensitivity, and adaptability to the dynamic network conditions inherent in 5G systems.

\subsubsection{Loss Function}
The autoencoder employs the mean squared error (MSE) as its loss function, which quantifies the reconstruction error over all time steps. The loss function is defined as:
\vspace{-5 pt}
\begin{equation}
L(x, \hat{x}) = \frac{1}{T} \sum_{t=1}^{T} \| x_t - \hat{x}_t \|^2
\vspace{-5 pt}
\end{equation}
where \( x \) represents the actual input sequence, and \( \hat{x} \) denotes the reconstructed sequence generated by the model.

\subsection{Long Short-Term Memory (LSTM) Units}
Our autoencoder model employs Long Short-Term Memory (LSTM) units to overcome the vanishing gradient problem encountered in traditional RNNs. Each LSTM unit is an assemblage of a cell that preserves values across varying time intervals, and a trio of gates—input, output, and forget—that manage the information flow within the unit.

The LSTM facilitates the preservation and regulation of information over time through intricate interactions of its state and gating mechanisms, which are mathematically characterized by the following equations:
\vspace{-5 pt}
\begin{align}
i_t &= \sigma(W_{ii}x_t + W_{hi}h_{t-1} + b_i) \notag \\
f_t &= \sigma(W_{if}x_t + W_{hf}h_{t-1} + b_f) \notag \\
o_t &= \sigma(W_{io}x_t + W_{ho}h_{t-1} + b_o) \notag \\
g_t &= \tanh(W_{ig}x_t + W_{hg}h_{t-1} + b_g) \notag \\
c_t &= f_t \odot c_{t-1} + i_t \odot g_t \notag \\
h_t &= o_t \odot \tanh(c_t)
\end{align}

In these equations, \( i \), \( f \), \( o \), and \( g \) denote the activation vectors for the input gate, forget gate, output gate, and cell input, respectively. The vector \( c_t \) represents the cell state, with \( W \) denoting the weight matrices corresponding to each gate, \( b \) as the bias vectors, \( \sigma \) as the sigmoid activation function, and \( \odot \) indicating the Hadamard product or element-wise multiplication.

In the context of our model's pseudocode, these LSTM units are instantiated within the encoding and decoding phases of the autoencoder, ensuring the retention and relevant transformation of temporal information. The LSTM's ability to selectively remember and forget information via its gates makes it particularly adept for sequence reconstruction in the presence of self-attention, which further augments the model's sensitivity to temporal dependencies and anomalies within 5G network traffic data.

Parameters such as \( W_{ii} \), \( W_{hi} \), 
and the bias vectors \( b_i \), \( b_f \), \( b_o \), \( b_g \) in equation 8 are encapsulated within the parameter set \( \Theta \) in the pseudocode. These parameters are optimized during the training procedure to minimize the reconstruction error as defined by the loss function \( L(x, \hat{x}) \) \cite{9030505}.

\subsection{Anomaly Detection Mechanism}
Anomaly detection in the proposed model is based on the reconstruction error. A threshold is set based on the error distribution of the training data, where instances with a reconstruction error above the threshold are classified as anomalies. The threshold is typically determined using a validation set or through cross-validation \cite{9744555}. The criterion for anomaly detection is defined as:
\vspace{-5 pt}
\begin{equation}
\text{Anomaly} = 
\begin{cases}
1, & \text{if } \text{error} > \text{threshold} \\
0, & \text{otherwise}
\end{cases}
\vspace{-5 pt}
\end{equation}

The proposed model provides a robust framework for unsupervised detection of anomalies in 5G networks. By leveraging the temporal correlation in I/Q samples and the LSTM's ability to remember long-term dependencies and self-attention capabilities, the model can effectively identify unusual patterns that signify potential security threats or network failures.

\section{Deployed Architecture, Metrics and Visualization}

The architecture of the model used for time-series intrusion detection in the context of 5G network traffic analysis is an ensemble of 5G testbed interfacing with the RNN-based autoencoder. This section will describe separately the two architectures of 5G testbed and autoencoder model. We will discuss the rationale behind the selection of various components, and the performance metrics used to evaluate the model's effectiveness.

\begin{figure}[t!] 
\centering
\includegraphics[width=\columnwidth]{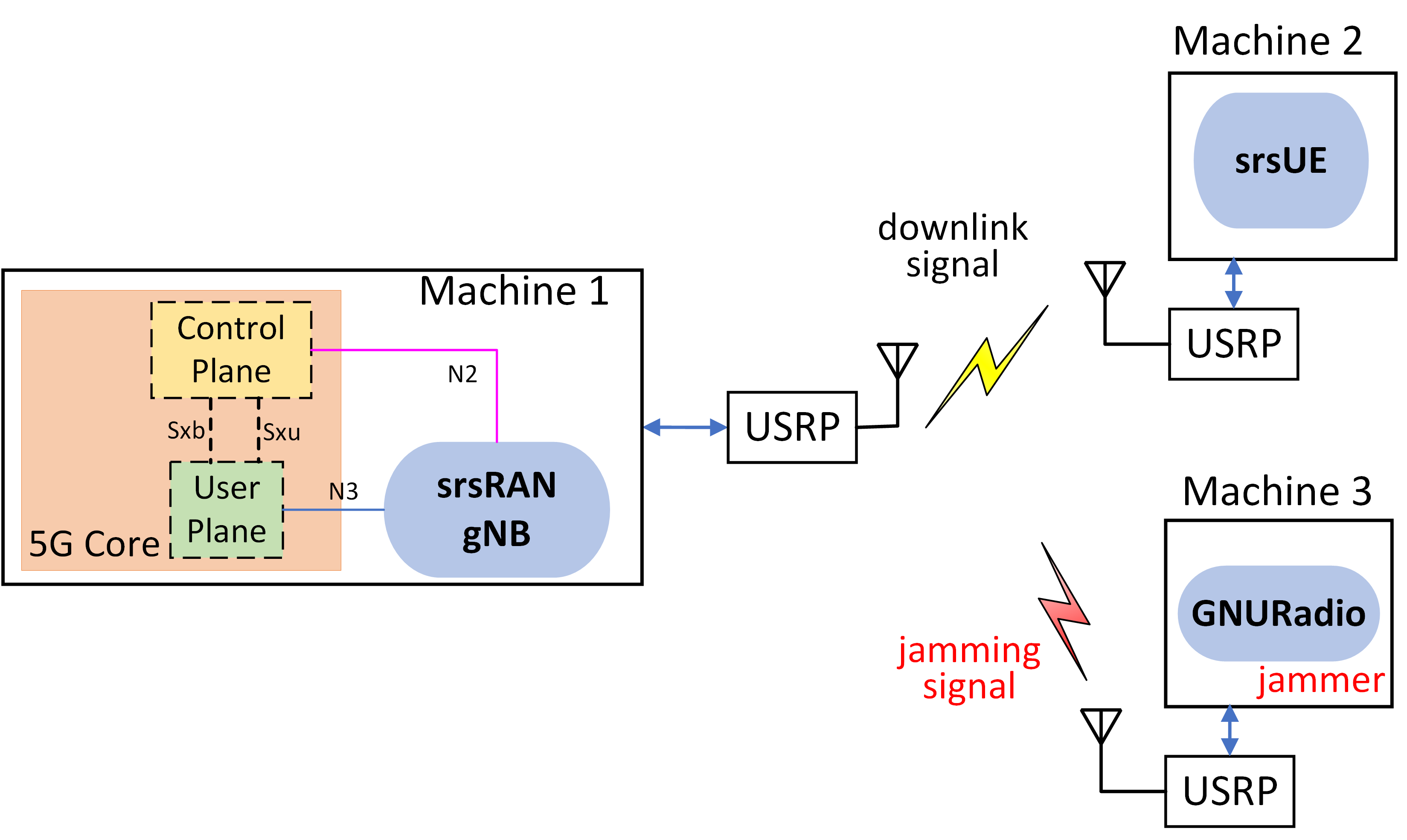}
\caption{5G network testbed and its architecture.}
\label{Testbed}
\vspace{-15pt}
\end{figure}
\subsection{\textbf{5G Network Testbed}}
Open-source software and software-defined radio (SDR) devices are used to build up a 5G network testbed, as shown in Fig. \ref{Testbed}. The srsRAN 
realizes both the gNB and UE, while open5GS fulfills all the functionalities of a 5G core network\cite{open5GS}. The gNB and the core network are co-located at \textit{Machine 1}. A UE is implemented in \textit{Machine 2}. The UE and gNB are connected via RF frontend devices called USRP and downlink signals are generated. At the same time, a jammer generates jamming signals to interfere the communications between gNB and UE by making use of GNURadio. The jamming signal's frequency, transmission power, bandwidth, pattern and modulation scheme are all configurable.   

In the established 5G testbed, the gNB sends downlink data to the legitimate UE, which a jammer attempts to interfere the downlink transmission by generating different kinds of jammin g signals with various frequency, bandwidth, and power. A RNN-based autoencoder is deployed at UE side to analyze the spectrum to identify if there is a jamming attack.     
\subsection{\textbf{Enhanced Autoencoder Model with Self-Attention in Encoder}}

The upgraded model architecture incorporates a significant enhancement through the integration of a self-attention layer, specifically within the encoder. This integration is a strategic decision aimed at bolstering the model's capability to process and interpret complex temporal sequences inherent in 5G network I/Q samples.

\subsubsection{Rationale for Self-Attention in Encoder}
The self-attention mechanism, is designed to augment the model's ability to analyze sequential input data. The encoder, being responsible for compressing the input into a latent representation, is the most critical phase for capturing the intricate dependencies and temporal relationships within the data. By implementing self-attention in this phase, the model gains an enhanced capacity to discern patterns and anomalies in the data that might otherwise be overlooked. The self-attention layer is strategically positioned after the initial LSTM layer in the encoder. This placement ensures that the raw sequential data is first processed by the LSTM to capture basic temporal relationships, which are then further refined by the self-attention mechanism. The multi-head attention framework within this layer enables the model to focus on different segments of the input sequence, assessing their relevance and contributing to a more accurate and nuanced latent representation.

\subsubsection{Hyperparameters Adjustment with Self-Attention Integration}
With the integration of the self-attention layer, it became imperative to reassess and finely tune the model's hyperparameters to achieve optimal performance. Table \ref{tab:model-architecture} encapsulates all the necessary details regarding the architecture and hyperparameters of the enhanced autoencoder model:
\vspace{-8 pt}
\begin{table}[ht]
\centering
\caption{Model Architectures and Hyperparameters}
\label{tab:model-architecture}
\begin{tabular}{|l|l|}
\hline
\textbf{Hyperparameter}                    & \textbf{Value}                                    \\ \hline
LSTM Units                            & 50/25                               \\ \hline
Self-Attention Layer                  & MultiHeadAttention Implemented              \\ \hline
LSTM Layers                & Two layers                            \\ \hline
Number of Attention Heads             & 4                                                 \\ \hline
Key Dimension                         & 50                                                \\ \hline
Activation Function                   & ReLU                                              \\ \hline
Optimizer                             & Adam                                              \\ \hline
Loss Function                         & MSE                         \\ \hline
Batch Size                            & 100/500/1024                                               \\ \hline
Sequence Length                       & 10/20/100         \\ \hline
Number of Epochs                      & 50/100/300                                                 \\ \hline
\end{tabular}
\vspace{-8 pt}
\end{table}


This enhanced model architecture, with the self-attention layer integrated into the encoder, represents a significant advancement in our ability to detect anomalies in 5G network I/Q samples. The iterative process of tuning the model's hyperparameters, particularly with the addition of self-attention, ensures a robust and efficient anomaly detection capability, tailored to the complexities of 5G network signal analysis.

\subsection{Visualization}
Visualization techniques are essential for the analysis and interpretation of the performance of anomaly detection models. A time series plot is one such technique, which juxtaposes the original and reconstructed I/Q data for a given sample, allowing discrepancies that may signal anomalies to be easily spotted. Histograms complement this by showing the distribution of reconstruction errors across the dataset, with a threshold line to mark the boundary between normal and anomalous readings. This helps in identifying the range within which most data points fall and the outliers that constitute potential anomalies. Additionally, threshold plots are particularly useful; by drawing a line at the chosen threshold level, they provide a clear visual indicator of the model's sensitivity to anomaly detection. This threshold is crucial for both setting the initial detection parameters and for ongoing adjustment to optimize the model's predictive accuracy. Together, these visualization tools form a comprehensive framework for evaluating the model's ability to detect anomalies and for fine-tuning its parameters for practical application.

\begin{figure}[htbp]
\centering
\includegraphics[width=0.99\columnwidth]{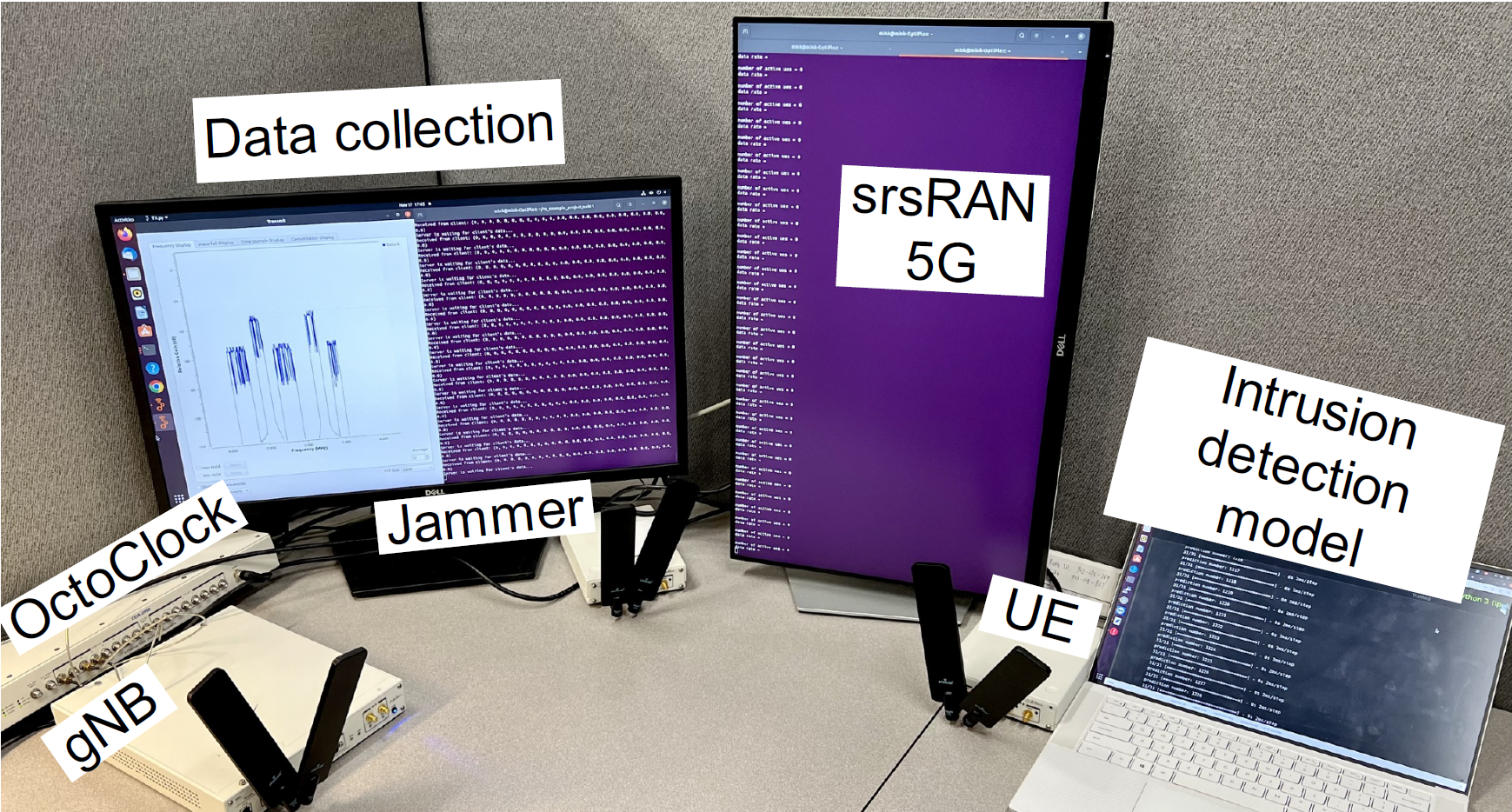}
\caption{Implementation 5G SDR-based testbed.}
\label{testbed_iq}
\end{figure}

\begin{figure*}[htbp]
    \centering
    \begin{subfigure}[b]{0.6\textwidth}
        \includegraphics[width=\linewidth]{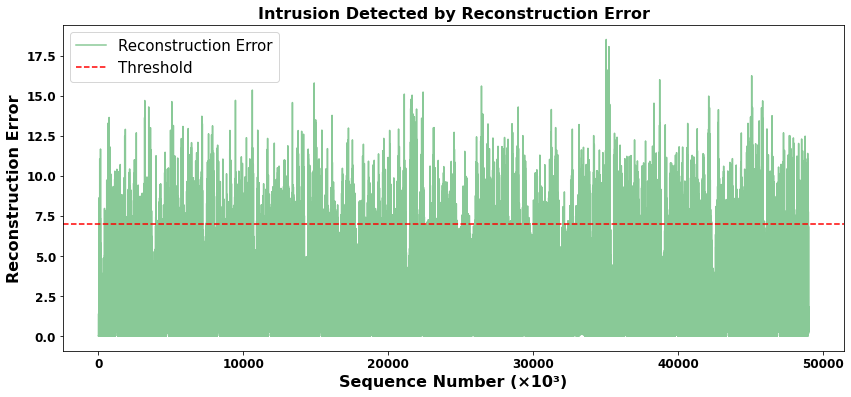}
        \caption{Reconstruction Error}
        \label{fig:image1}
    \end{subfigure}
    \hfill
    \begin{subfigure}[b]{0.49\textwidth}
        \includegraphics[width=\linewidth]{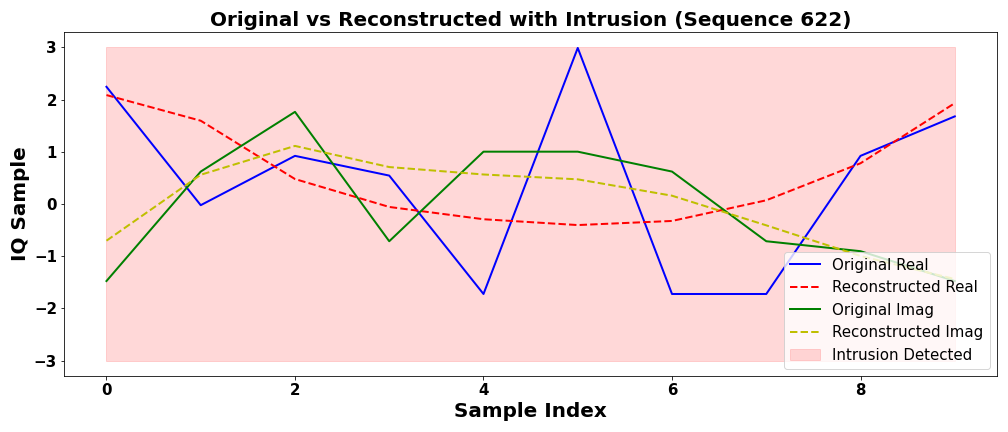}
        \caption{Illustration of Intrusion Detection Sequence}
        \label{fig:image2}
    \end{subfigure}    
    \begin{subfigure}[b]{0.49\textwidth}
        \includegraphics[width=1.0\linewidth]{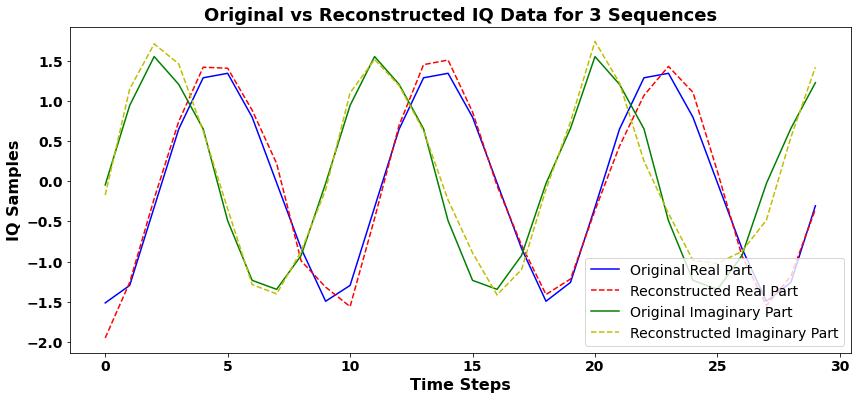}
        \caption{Normal Output: Original vs. Reconstructed IQ Data}
        \label{fig:image3}
    \end{subfigure}
    
    \caption{{Figures show the experiment results of reconstructed errors (a), detected sequence of intrusions (b), and the normal output of the system (c).}
    }
    \label{results}
\vspace{-15pt}    
\end{figure*}

\section{Implementation and Test Results}

To establish a robust experimental foundation for intrusion detection in 5G networks, we systematically executed a series of procedures encompassing scenario development, data collection and processing, model architecture selection, and training, followed by the tuning of models and hyperparameters. We then proceeded to the visualization of the model's performance. This methodical approach ensures that each phase of the experiment contributes to a robust validation of the proposed intrusion detection system. We present the details of these procedures as follows.

\subsection{Scenario Development}
Our research commenced with the development of realistic scenarios that mirror potential adversarial attacks within a 5G network. By designing these scenarios, we aim to create a diverse set of conditions under which the robustness and adaptability of our detection models can be rigorously evaluated. Our attack scenario is shown in Figure \ref{Testbed}.

\subsection{Data Collection and Preparation}
Data collection was meticulously carried out to capture a comprehensive range of normal operational patterns from an actively deployed 5G network. Our dataset, now enriched to include approximately 300 million I/Q samples from UE-gNodeB communications, serves as the foundational bedrock for anomaly detection learning. The construction of this dataset was pivotal in mirroring the intricate operational state of a live 5G network, which is essential for the subsequent training of our detection models.

For the purpose of testing and validation, adversarial elements were synthetically infused into this dataset to simulate a spectrum of jamming attacks. These attacks, both static and dynamic in nature, were introduced to test the resilience and detection capabilities of our system under realistic interference conditions. The experimental setup, as illustrated in Figure \ref{testbed_iq}, showcases the implemented 5G network configuration complete with the jammer in operation. This setup not only facilitated the generation of a robust dataset reflective of actual network conditions under duress but also enabled the precise tuning of our intrusion detection model to identify and respond to such adversarial tactics effectively.

The figure captures our practical testbed environment where the srsRAN 5G tool is interfacing with both the gNodeB and UE, as well as the OctoClock and jamming devices. This comprehensive setup is indicative of our approach to data collection—deliberate and grounded in real-world applicability, ensuring that the collected data underpins the veracity of our intrusion detection system's operational readiness.
\vspace{-6 pt}


\subsection{Model Architecture}
In exploring suitable model architectures for our autoencoder, we considered LSTM-based configurations for their capability to capture temporal dependencies in sequential data. Two configurations were evaluated: a baseline model with a symmetrical layout of LSTM layers, and a complex model with a higher initial number of units to investigate the trade-offs between model capacity and overfitting potential.

\subsection{Model Training, Evaluation, and Hyperparameter Tuning}
Model training was conducted with an emphasis on iterative improvement and overfitting prevention. We employed a range of hyperparameters, systematically adjusted to optimize the training process. Our regimen included extensive cross-validation to refine these parameters, ensuring the model's generalization capabilities across unseen data.
Post-training, we leveraged visualization to interpret the model's performance. This entailed generating detection matrices and error characteristics plots to provide a multi-faceted understanding of the intrusion detection model's capabilities and limitations. The details of the experiment scenarios and the corresponding configurations are presented in the following table:

\subsection{Test Results and Discussion}

This section will provide a detailed discussion on the test results, including an analysis of the models' intrusion detection performance and the implications of our findings for future 5G security protocols. Upon illustrating the test results in Figure \ref{results}, several observations emerge regarding the intrusion detection capabilities of the proposed model in a 5G network context. The reconstruction error plot serves as an immediate visual reference for identifying sequences that surpass the defined threshold for anomaly detection. This threshold demarcation allows us to pinpoint the exact instances where the autoencoder perceives irregularities, which can be instrumental in tuning the model for better false positive rates. This visualization not only confirms the efficacy of the threshold value, but also sheds light on the error distribution across the sequence of I/Q samples.

In the first figure, the reconstruction error across numerous sequences is displayed. The intermittent spikes beyond the threshold level effectively signal the presence of anomalies, validating the model's ability to detect intrusions across a large dataset. The threshold setting plays a crucial role here, balancing the sensitivity between false positives and missed detections.

Comparing original and reconstructed I/Q samples through time series plots allows for a direct visual assessment of the model's reconstruction fidelity. Discrepancies between the original and reconstructed signals may indicate potential anomalies or model shortcomings in capturing the signal complexity. Such plots are invaluable for iterating on model architecture and training regimens to enhance performance.

In this regard, the second figure presents a specific instance of this comparison where the model detected an intrusion. It is evident that the reconstructed values diverge significantly from the original I/Q samples when the intrusion occurs, as highlighted by the shaded area. This detection is aligned with the predefined threshold, confirming the model's effective identification of abnormalities within the signal.

The third figure illustrates the comparison between original and reconstructed I/Q data in normal sequences. The close alignment between the original and reconstructed values for both real and imaginary components indicates the model's proficiency in learning the normal operational patterns. 

\section{Conclusion}
The results not only confirm the efficacy of the attention-enhanced RNN autoencoder in detecting anomalies in 5G network traffic, but also highlight the model's potential in differentiating between normal and adversarial patterns with high accuracy. The implementation of this model within real-world 5G networks could significantly bolster security measures against a wide array of potential attacks.

Further research could expand on the dataset variability and explore the model's resilience against more sophisticated adversarial tactics. Additionally, the impact of different attention mechanisms and their configurations on model performance warrants a deeper investigation to fine-tune the system for operational deployment.

\section*{\textcolor{black}{Acknowledgment}}
\noindent
This work is supported by NSF and the Office of the Under Secretary of Defense (OUSD) – Research and Engineering, under Grant ITE2326898, as part of the NSF Convergence Accelerator Track G: Securely Operating Through 5G Infrastructure Program. The work of Mohammadreza Kouchaki, Minglong Zhang, Aly S. Abdalla, and Vuk Marojevic is partially supported by the Office of Naval Research under Award No. N00014-23-1-2808.

\balance


\input{main.bbl}

\end{document}